\documentclass{raa}            
\usepackage{graphicx,times}      
\usepackage{natbib}

\providecommand{\bm}[1]{\mbox{\boldmath$#1$\unboldmath}}

\begin{document}

   \title{Recovering the Pulse Profiles and Polarization Position Angles of Some Pulsars from Interstellar Scattering
 $^*$
\footnotetext{\small $^{*}$ Supported by the National Natural Science Foundation of China.}
}

 \volnopage{ {\bf 2011} Vol.\ {\bf 9} No. {\bf XX}, 000--000}
   \setcounter{page}{1}

   \author{Abdujappar Rusul
      \inst{1,3}
   \and Ali Esamdin
      \inst{2,3}
   \and Alim Kerim
      \inst{1,3}
   \and Dilnur Abdurixit
      \inst{1,3}
   \and Hongguang Wang
       \inst{4}
   \and Xiao-Ping Zheng
       \inst{5}
   }

   \institute{School of Physics Science and Technology, Xinjiang University, Urumqi
   830046, China
        \and
             Xinjiang Astronomical Observatory, Chinese Academy of Science, Urumqi 830011, China; {\it aliyi@xao.ac.cn}\\
        \and
             Xinjiang University-National Astronomical Observatory Joint Center for Astrophysics, Urumqi 830046, China
        \and Center for Astrophysics, Guangzhou University, Guangzhou 510006, China\\
        \and The Institute of Astrophysics, Huazhong Normal University, Wuhan 430079, China\\
\vs
   {\small Received [year] [month] [day]; accepted [year] [month] [day] }
}

\abstract{ Interstellar scattering causes broadening and distortion to the mean pulse profiles and polarization position angle (PPA) curves, especially to the pulse profiles observed at lower frequency. This paper has implemented a method to recover the pulse profiles and the PPA curves of five pulsars which have obvious scattered pulse profiles at lower frequency. It reports a simulation to show the scattering and descattering of pulse profiles and PPA curves. As a practical application, lower-frequency profiles and PPA curves of PSR 1356-60, PSR 1831-03, PSR 1838+04, PSR 1859+03, PSR 1946+35 are obtained. It is found that the original pulse profiles and PPA curves can be recovered.
}

   \authorrunning{Abdujappar Rusul, Ali Esamdin, Alimjan \& Dilnur }            
   \titlerunning{Recovering the Pulse Profiles and PPA curvse of Some Pulsars from Interstellar Scattering}
   \maketitle


%
%
\section{Introduction}           
\label{sect:intro}

Interstellar medium (ISM) scattering broadens the intrinsic lower frequency pulse profiles of pulsar and causes flattening and distortion of the PPA curves (\cite{Li03}; \cite{Karastergiou09}) to some extent according to the observed frequency and the distribution scale of ISM which is located between pulsar and observer. The scattering of pulse profiles has been studied extensively since the first observation of scintillation of pulsars (\cite{scheuer68}). Since then, pulsar researchers developed a several ISM scattering model which were the  thin screen model (\cite{rankin70}; \cite{komersaroff72}),  the thick screen and the extended screen model (\cite{wiliamson72}), based on the observable effects of temporal broadening of  pulse profiles and assumable scale of a scattering screen in the ISM. The scattering effect of the pulse broadening and the flattening of PPA curves are studied frequently by many authors (\cite{komersaroff72}; \cite{Rickett77}; \cite{Li03} etc) and the distortion of PPA curves with orthogonal jumps by \cite{Karastergiou09}. They used the higher frequency mean pulse profile without obvious scattering as the intrinsic pulse profile and convolved it with scattering models for obtaining similar pulse-shapes and PPA curves as observed scattered lower-frequency pulse profiles. Only a few of them have performed deconvolution to recover total intensity pulse profiles of lower frequency from scattering (\cite{Weisberg90}; \cite{Kuzmin93}; \cite{Bhat03}). This paper revisited the method of \cite{Kuzmin93} to restore total intensity pulse profiles of pulsar and extended it to the restoration of linear intensity profiles and PPA curves for another five pulsars. Scattering broadening time scales of those pulsars has been obtained from best fit for three different scattering models. The scattering time scale is a key parameter in all scattering models, which depends on observing frequency and dispersion measure (DM) (\cite{Ramachandran97}).

       Descattering compensation for the first Stokes parameter $I(t)$ of the scattered pulse signal was performed by \cite{Kuzmin93}; their method worked well for recovering the original low-frequency pulse profiles of the crab pulsar, but when discussing the restoration of the rest of the Stokes parameters, it draws our consideration whether all the Stokes parameters scattered the same way as $I(t)$. In early works of \cite{komersaroff72} and \cite{Rickett77}, they assumed that the scattering effect may be approximated by convolving each of the time-dependent Stokes parameters of unscattered pulse with scattering model under certain assumptions. In the research note of \cite{Li03}, based on the works of \cite{Macquart00} they simply assumed that scattering process works similarly on all Stokes parameters. By using that assumptions in their convolution method they explained well the scattering effect on pulse broadening and on PPA curve flattening, but their approach dose not work properly when applied in the deconvolution method of \cite{Kuzmin93} to recover the shape of PPA curve. This research paper uses the same method as \cite{Kuzmin93} to recover the total intensity profile $I(t)$; to recover the linear intensity and the PPA curve, this paper assumes that the complex number form of Stokes parameters $Q, U$ may scatter the same way as $I(t)$; Stokes parameters $Q$ and $U$ has been treated as the real and the imaginary component of a complex number respectively (\cite{Jaap96}); such treatment is also applied in PSRCHIVE\footnotemark in the section of Complex-valued Rotating Vector Model; these assumption and method are applied to recover the pulse intensity profiles and PPA curves of some pulsars in section 3; results are discussed in section 4 and the conclusions are presented in section 5.

\footnotetext{ http://psrchive.sourceforge.net/manuals/psrmodel/.}


\section{Scattering models and method}
\label{sect:Obs}

This paper tests all three different kinds of scattering models to check the scattering phenomena in pulse profiles and in PPA curves. The thin screen model (Eq. 1) in which the signal is assumed to be scattered approximately mid-way between source and observer by irregularities in the ISM; the thick screen model (Eq. 2) in which ISM irregularities are distributed on a larger scale than thin screen model and it can be near the observer or near the source; the third one is the extended screen model (Eq. 3) in which the signal is scattered in the whole path of its propagation, so that the irregularities spread in the whole space between the source and the observer (\cite{wiliamson72}). The functions of those models are following:

\begin{equation}
g_{thin}=\exp(-t/\tau_{s}) ~~~~~~~~~~~~~~~~~~~~~~~~~~~~~~~~~(t \geq 0)
\end{equation}

\begin{equation}
g_{thick}=(\frac{\pi\tau_{s}}{4t^3})^{1/2}\exp(\frac{-\pi^2\tau_{s}}{16t}) ~~~~~~~~~~~~~~~(t>0)
\end{equation}

\begin{equation}
g_{extend}=(\frac{\pi^5\tau_{s}^3}{8t^5})^{1/2}\exp(\frac{-\pi^2\tau_{s}}{4t}) ~~~~~~~~~~~(t>0)
\end{equation}
$\tau_{s}$ is scattering broadening time scale which can be determined through an empirical relation between wavelength ($\lambda$) and DM (\cite{Ramachandran97})

\begin{equation}
\tau_{s}= 4.5\times10^{-5}DM^{1.6}\times(1+3.1\times10^{-5}\times DM^{3})\times\lambda^{4.4}
\end{equation}
this relation is used as a reference to set an upper limit in the process of best fit, except for PSR B1946+35.

This paper has employed the method of compensation of \cite{Kuzmin93} with that of three different scattering models. His method works efficiently to recover the original shape of pulse profiles $x(t)$ from observed pulse profiles $y(t)$; here, the same method has been used to recover total intensity pulse profiles, the recovering of the linear intensity and PPA curve will be introduced at the end of this section. When $g(t)$ is assumed to be a scattering model or response function, the observed $y(t)$ is the product of $x(t)$ and $g(t)$, and the spectrum of recovered pulse (original pulse) can be written as (\cite{Kuzmin93})
\begin{equation}
X(f)=Y(f)/G(f)                                   \\
\end{equation}

\begin{equation}
Y(f)=\int y(t)\exp(-j2\pi ft)dt
\end{equation}

\begin{equation}
G(f)=\int g(t)\exp(-j2\pi ft)dt
\end{equation}
$Y(f)$ is the spectrum of observed pulse, $G(f)$ is the frequency response of scattering screen, and the descattered restored pulse $x(t)$
is obtained by inverse Fourier transformation:
\begin{equation}
x(t)=\int X(f)\exp(j2\pi ft)df
\end{equation}
by using the above equations, the total intensity pulse profiles can be recovered from the scattered pulse profiles $y(t)$.
It is also possible to study the effect of scattering on pulse profiles and on PPA curves by changing the descattering procedure above as Eq.(9) and Eq.(10). The author has tried the work of \cite{Li03} by using Eq.(9)and Eq.(10) with complex treatment of Stokes parameters $Q, U$. And obtained the same results in explaining pulse profile broadening and PPA curve flattening caused by ISM scattering. The followings are modified functions of \cite{Kuzmin93} to hold convolution for repeating the work of \cite{Li03}. The spectrum of the scattered pulse is calculated by
\begin{equation}
Y(f)=X(f)*G(f)
\end{equation}
and the scattered pulse $y(t)$ would be written
\begin{equation}
y(t)=\int Y(f)\exp(j2\pi ft)df
\end{equation}

  The above equations (5$-$8) were used to recover total intensity profiles of the crab pulsar from ISM scattering (\cite{Kuzmin93}). This time they are used to recover total intensity, linear intensity profiles and PPA curves of a few other pulsars. In the process of restoration of Stokes parameters of $Q, U$, it is assumed that the observed scattered Stokes parameter $Q, U$ form a complex number $z(t)=Q(t)+iU(t)$; $p(t)$ is descattered complex number from $z(t)$. The observed spectrum of $z(t)$ would be
\begin{equation}
Z(f)=\int z(t)\exp(-j2\pi ft)dt
\end{equation}
  The spectrum of a descattered $p(t)$ is
\begin{equation}
P(f)=Z(f)/G(f)
\end{equation}
so, the descattered recovered linear intensity and PPA can be obtained from
\begin{equation}
p(t)=\int P(f)\exp(j2\pi ft)df
\end{equation}
complex treatment to the Stokes parameters $Q, U$ is more practical than treating them separately as scalar; when the $Q, U$ is tested as scalar separately by using the method of recovering of $I(t)$, the results in all three models for descattering of PPA curves are not as expected, and fail to produce smooth Swing-curves (\cite{Radhakrishnan69}), PPA jumps, smooth flat curves or to show similarity with the higher frequency pulse's PPA curves, but when the $Q, U$ are expressed as a vector over complex plane and using the method of \cite{Kuzmin93} it produces better results, the linear intensity and PPA curve are recovered more desirably.


\section{Simulation and practical application}
\label{sect:data}


\subsection{Simulation of scattering and descattering of pulse profiles and PPA curves}

This paper has held a simple simulation on intensity profiles and on PPA curves with three different models for scattering and descattering; as an example, the thick screen model is used. The simulated pulse profiles have a Gaussian shape, with PPA($\psi$) curves following the Rotating-vector model (RVM) (\cite{Radhakrishnan69}). This paper has assumed a coherent radiation of 100$\%$ polarized in which the degree of linear polarization in total intensity is 0.8, so that the Stokes parameters $Q, U$ can be expressed in terms of the degree of linear polarization and PPA by $Q=0.8Icos(2\psi)$, $U=0.8Isin(2\psi)$ (\cite{van57}). The simulated plots from (a) to (k) are shown in Figure 1; the left panel plots of (a), (b), (g), (h) are the original pulse profiles and PPA curves, the middle panel figures (c), (d), (i), (j) are the scattered pulse profiles and PPA curves, the right panel figures (e), (f), (k), (l) are the descattered restored pulse profiles and PPA curves; the solid lines and dotted lines in normalized intensity profiles (a), (c), (e), (g), (i), (k) are total and linear intensity profiles respectively; the dotted lines in the plots (b), (d), (f), (h), (j), (l) are PPA curves.  It can be seen from the figure of the simulation, the scattering dose caused pulse broadening and PPA curve flattening, and if the scattering screen $g(t)$ is definite it is in principle possible to recover the actual features of emitting signal.

   \begin{figure}
   \centering
   \includegraphics[width=3.091cm, angle=-90]{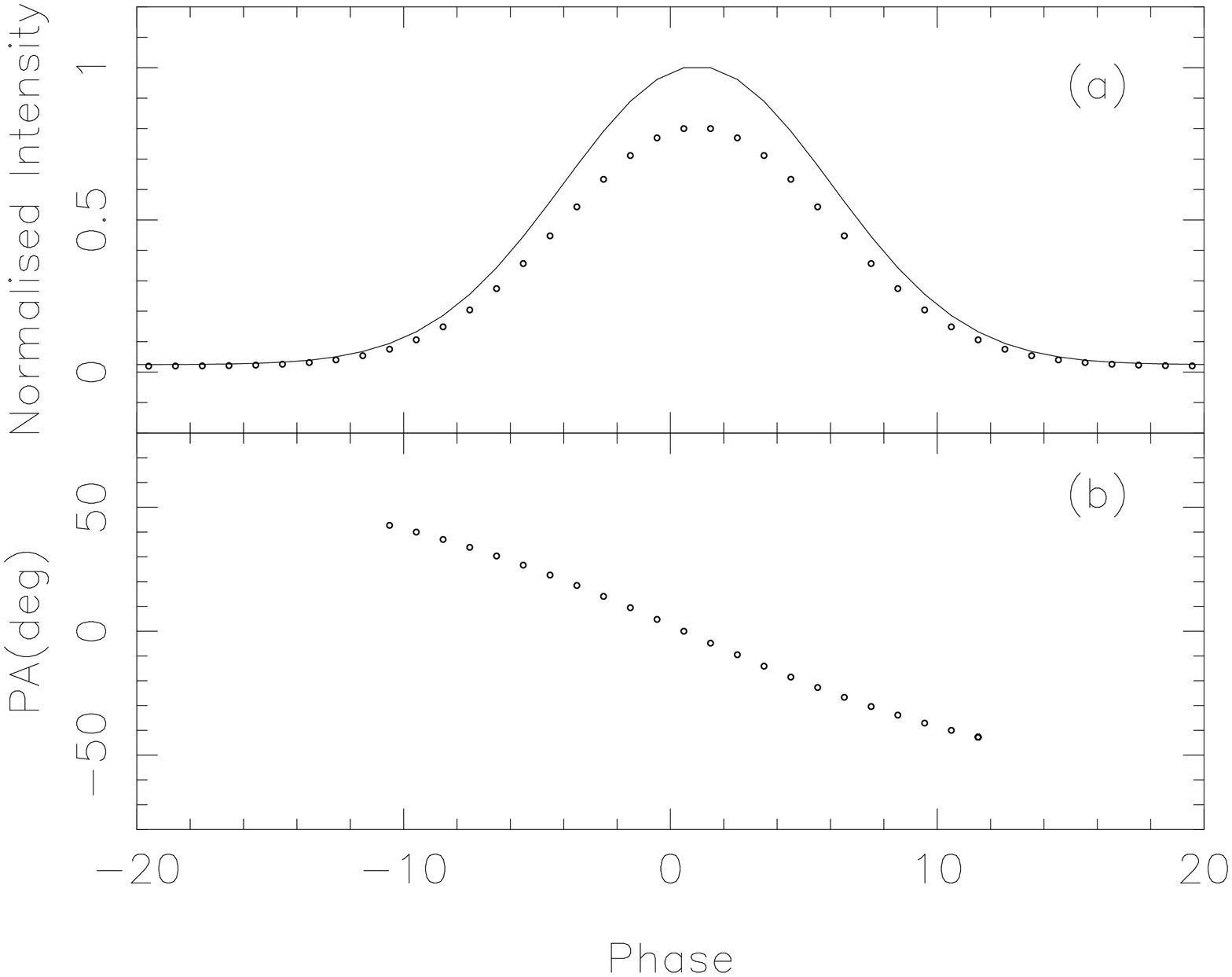}
   \includegraphics[width=3.090cm, angle=-90]{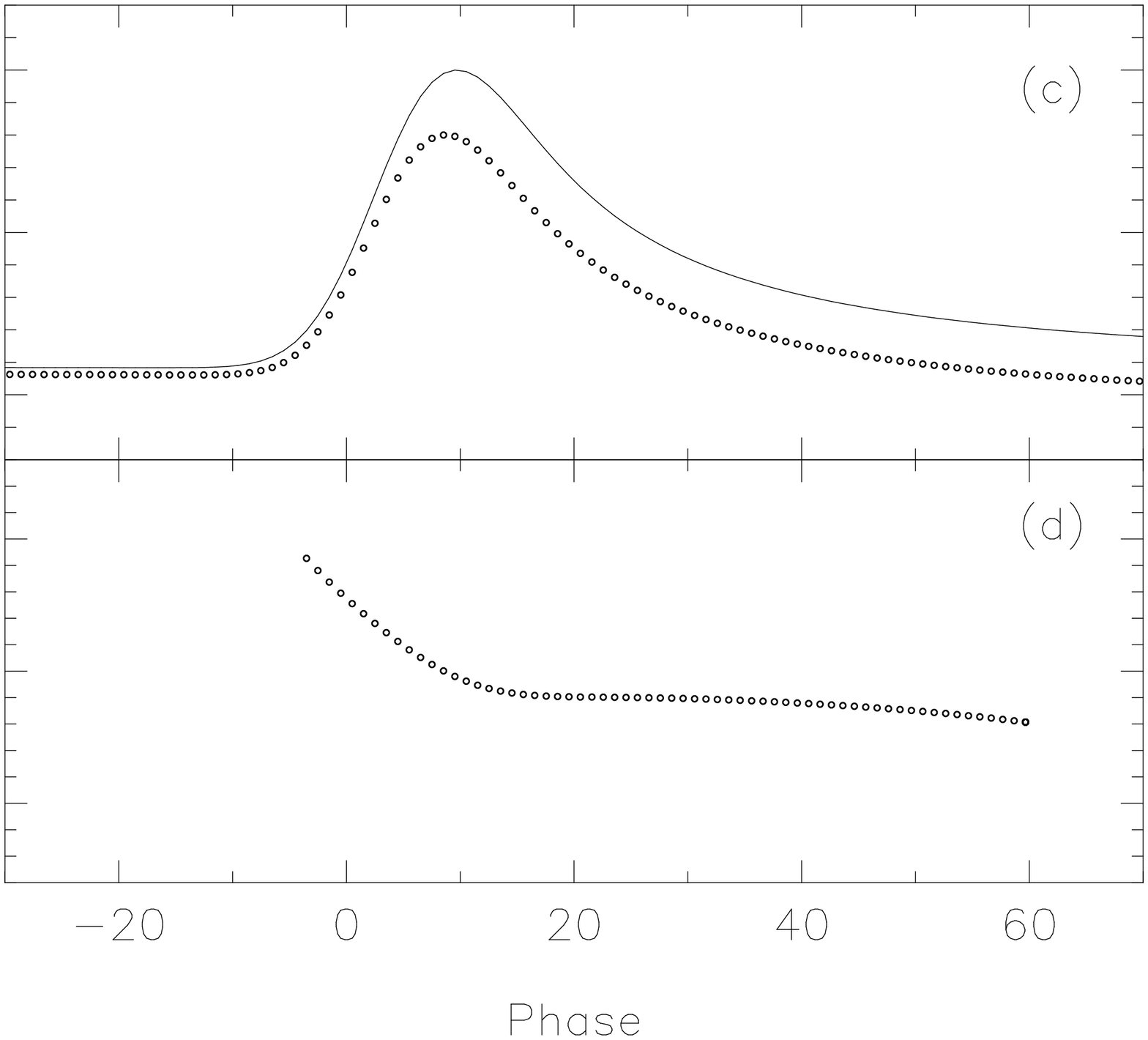}
   \includegraphics[width=3.10cm, angle=-90]{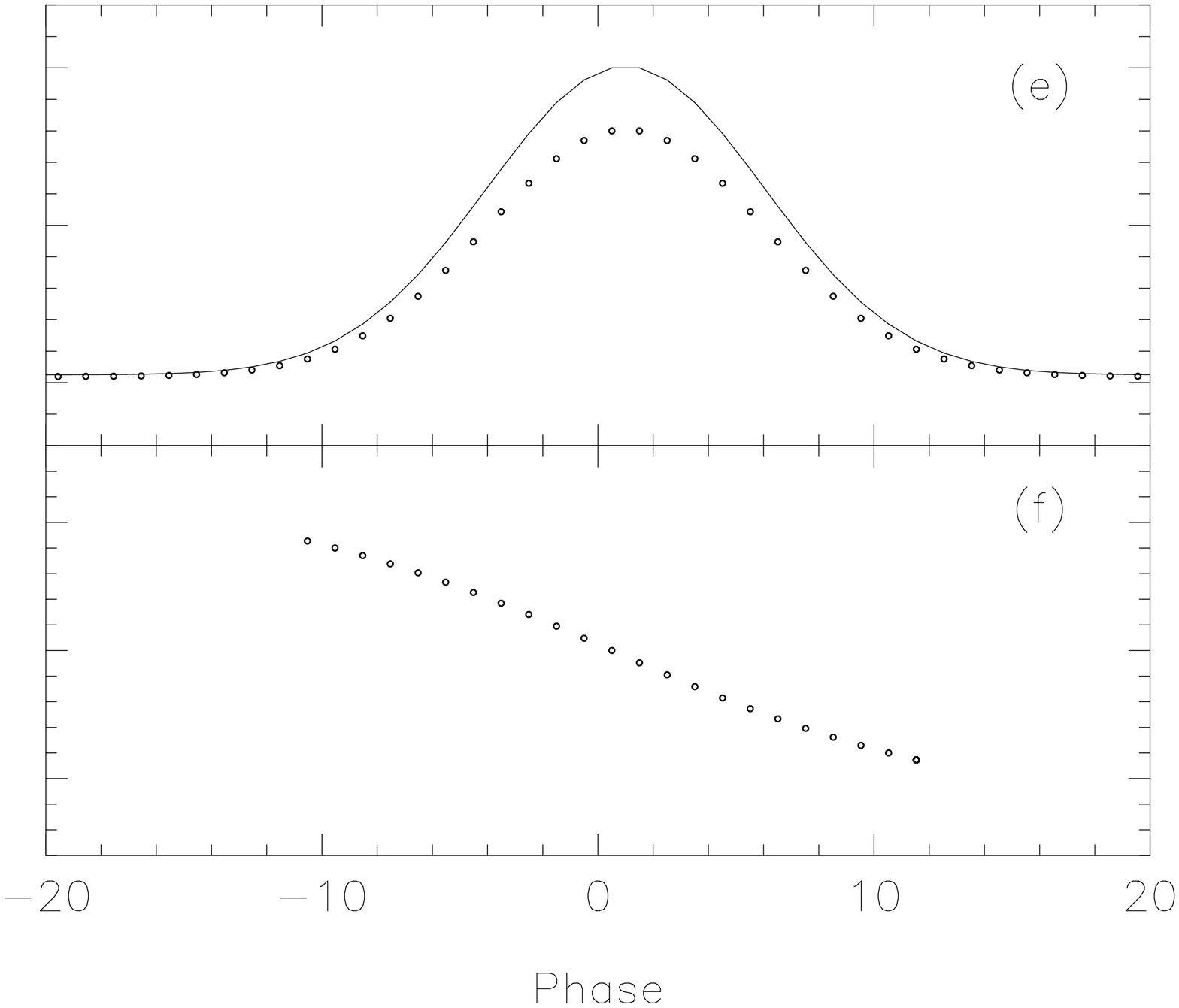}
   \includegraphics[width=3.0910cm, angle=-90]{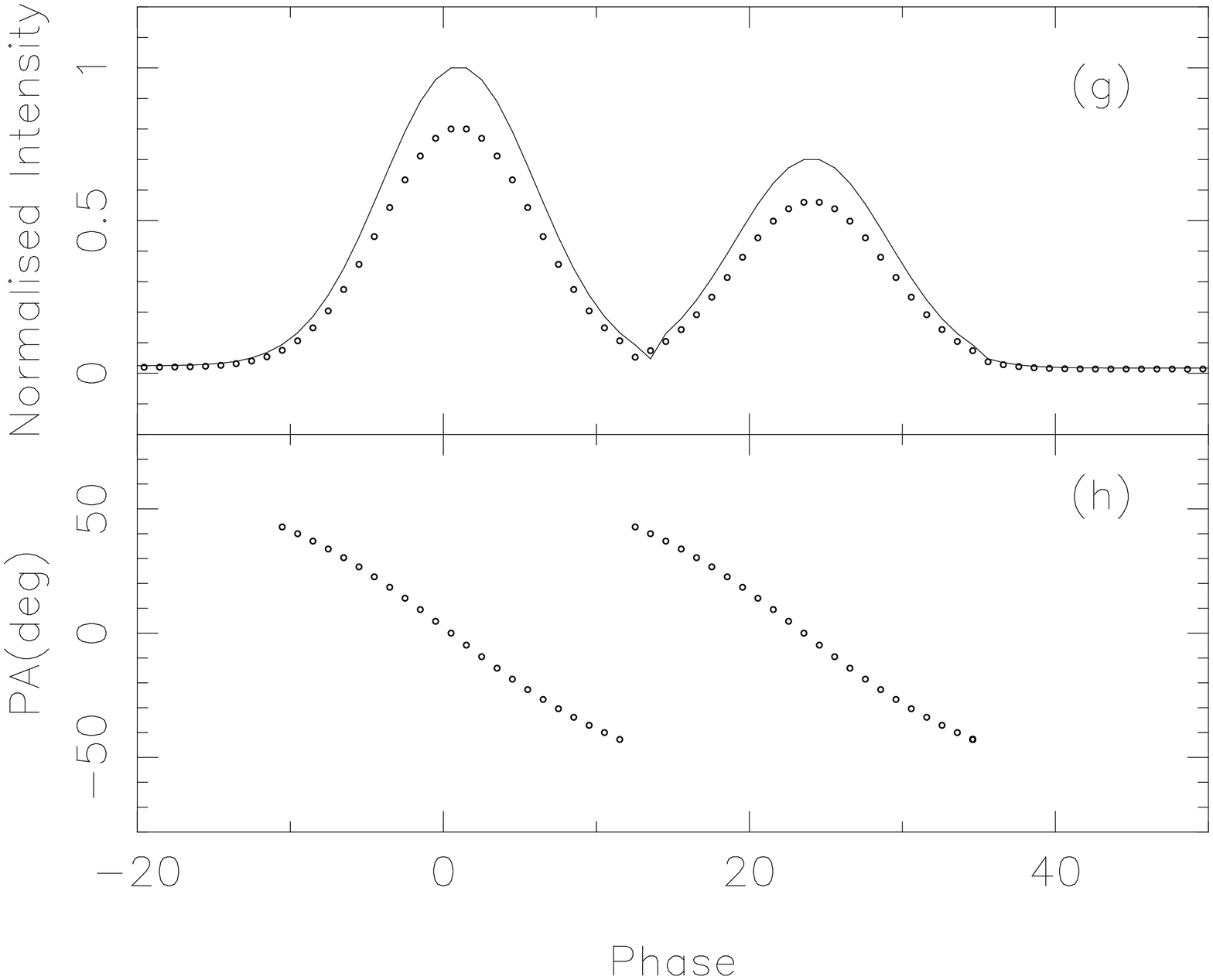}
   \includegraphics[width=3.08cm, angle=-90]{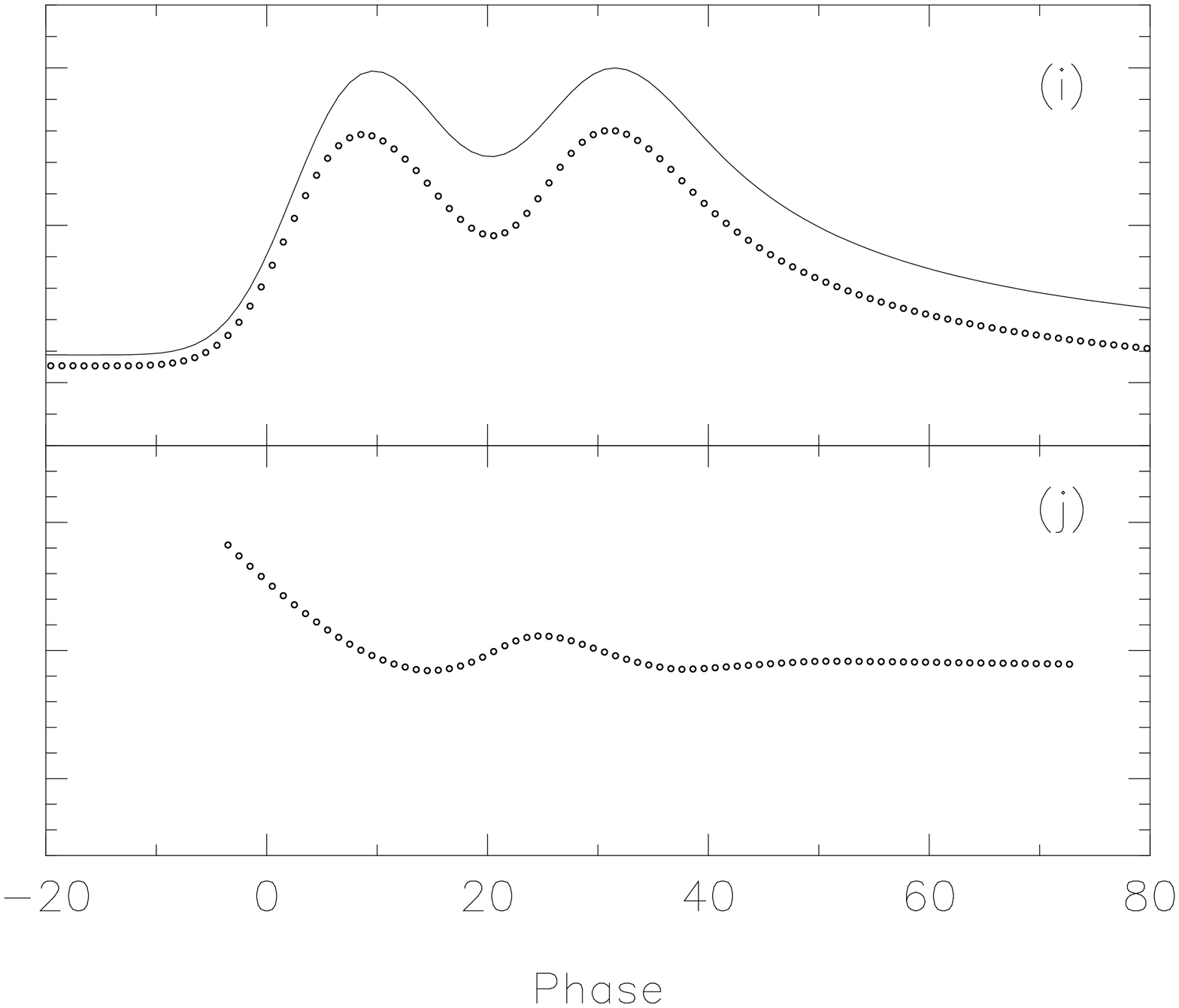}
   \includegraphics[width=3.110cm, angle=-90]{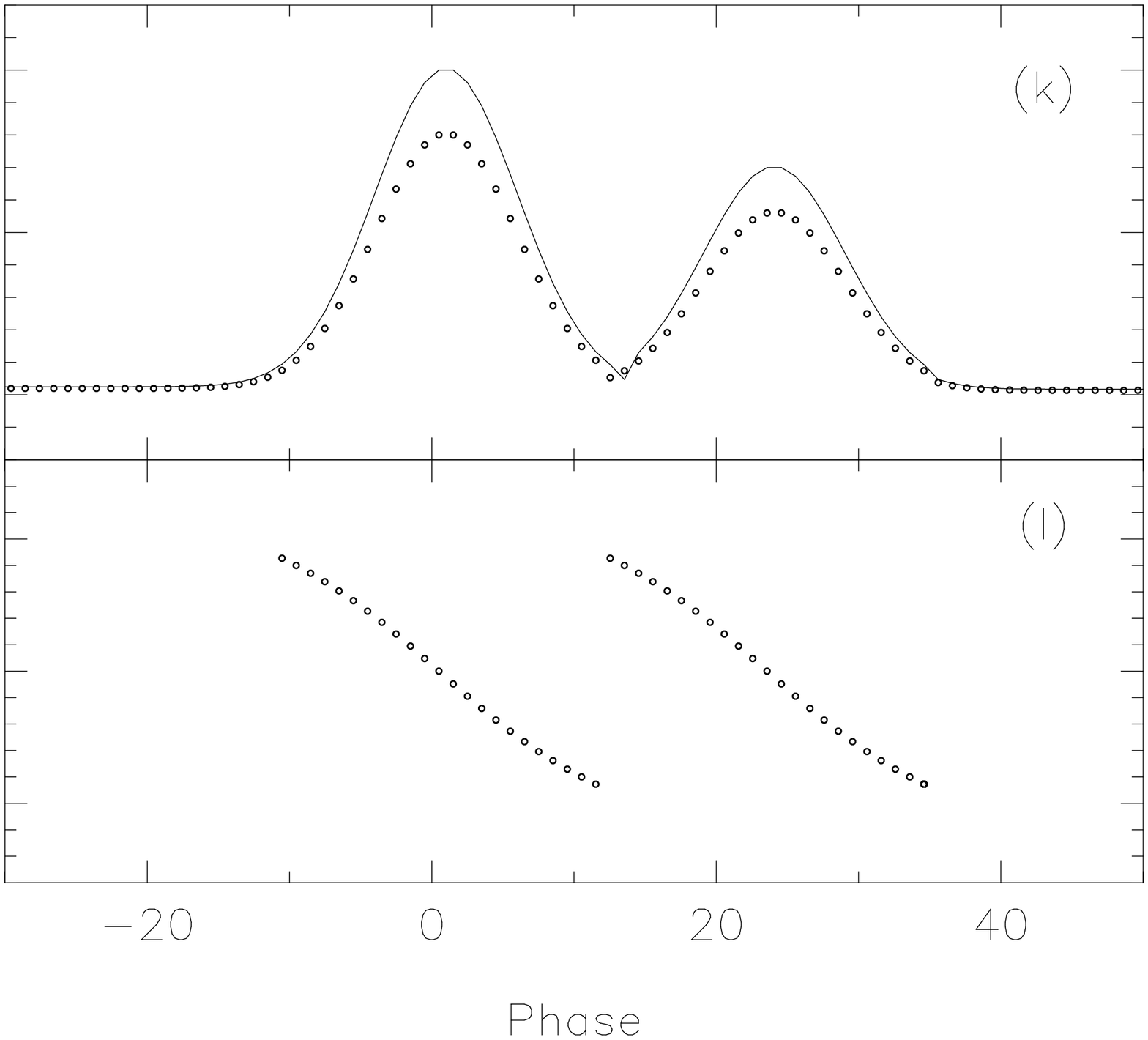}
   \begin{minipage}[]{125mm}

   \caption{ Pulsar signal simulation for scattering and descattering. the left panel plots of (a), (b), (g) and (h) are original pulse profiles and PPA curves, the middle panel plots of (c), (d), (i) and (j) are scattered pulse profiles and PPA curves and the left panel plots of (e), (f), (k) and (l) are  descattered pulse profiles and PPA curves. In all intensity pulse profiles, solid lines are total intensity and dotted lines are linear intensity; the dotted lines in plots of (b), (d), (f), (h), (j), (l) are PPA curves. } \end{minipage}
   \label{Fig1}
   \end{figure}

\subsection{Recovering the intensity pulse profiles and PPA curves  }
To compensate the scattered pulse profiles and PPA curves, the author of this paper searched in European Pulsar Network (EPN) online database \footnotemark (\cite{Lorimer98}) for sample pulsars (\cite{Gould98}) which have obvious pulse profile broadening; five pulsars' data are downloaded, four of which were previously used by \cite{Li03} for studying the effect of scattering on pulse profiles and on PPA curves. This study's calculation has used the lower-frequency pulse profiles with obvious scattering to be compensated for scattering; in all figures from Fig. 2 to Fig.6, the higher frequency profiles without obvious scattering (intrinsic pulse profiles) and it's PPA curves (intrinsic PPA curves) are given for comparison. The scattering time scale and some relative data of five pulsars are given in Table 1; the three different time scales for three different scattering models are obtained by the best fit for the pulse peak (Col.7, Col.8, Col.9), the time scales in Col.6 are calculated by Eq.(4) (see Table 1). When holding the best fit, almost in all pulsar the precision of those scattering time scales is approximately controlled in 1$ms$.

\footnotetext{ http://www.mpifr-bonn.mpg.de/old$_{-}$mpifr/div/pulsar/data/browser.html}



\begin{table}

\bc

\begin{minipage}[h]{146mm}

\caption[h]{ Parameters of five pulsars and their scattering time scales for different scattering models. The parameters are tabulated from column one to column nine as pulsar name, period, dispersion measure, observed higher frequency and lower frequency, empirical value of scattering time scale by Eq. (4), time scale for thin-screen, time scale for thick-screen, time scale for extended screen.}\end{minipage}

\small
 \begin{tabular}{ccccccccccc}
  \hline\noalign{\smallskip}
PSR Name & $P$ & $DM$ &  $Freq$  &$Freq$  &$\tau_{em}$  &$\tau_{thin}$  &$\tau_{thick}$  &$\tau_{extend}$ \\
   &$(ms)$  &$(pc$ $cm^{-3})$  &$(GHz)$  &$(GHz)$  &$(ms)$  &$(ms)$   &$(ms)$   &$(ms)$      \\

  \hline\noalign{\smallskip}
B1356$-$60   &  127.503335  & 294.133   &1.56     &0.659594    &9.88     &1.0     &0.5  &0.6\\
B1831$-$03   &  686.676816  &  235.800  &0.610    &0.408   &29.63     &15.0     &10.0    &6.5\\
B1838$-$04   &  186.145156  & 324.000   &0.925    &0.606    &22.38     &15.0     &3.0   &1.0\\
B1859$+$03   &  655.445115  & 402.900   &0.925    &0.606   & 60.97    &13.0     &7.4  &5.5\\
B1946$+$35   &  717.306765  &  129.050  &0.61    &0.408    &1.87     &10.0    &6.0     &4.0    \\

  \noalign{\smallskip}\hline
\end{tabular}
 \ec
\end{table}

Actually, because of the uncertainty of the scattering screen, any of the scattering models are not strictly appropriate to explain the scattering effect on pulse signals independently, even if we take  the pulse width evolution with observing frequency into account (\cite{lorimer05}). In general, this study has tested all three different scattering models to recover the pulse profiles and PPA curves; as an example, Fig. 2 to Fig. 6 presented the application of one of the three models for restoring pulse profiles and PPA curves; the intensity profiles are normalized to its own peak of intensity; the error of the PPA calculated the same with \cite{von97}by
\begin{equation}
\triangle \psi = \frac{\sqrt{(Q\cdot rms_{U})^2+(U\cdot rms_{Q})^2}}{(2L^2)}
\end{equation}
in all Figures from 2$-$6 the solid lines in the plots are the recovered total intensity (c) and linear intensity profiles (d), the dotted lines in (a), (b) are scattered total intensity profiles (upper panel) and linear intensity profiles (lower panel), the dotted lines in (c),(d) are total intensity (upper panel) and linear intensity (lower panel) profiles of intrinsic pulse profiles for comparison, the plots on the right are scattered PPA curves (upper panel (e)), PPA curves of intrinsic pulse signals (upper panel (f)) and recovered PPA curves (lower panel (g)). $"5\triangle \psi"$ error bars are presented in all Figures of plots of (e), (f), (g).

For PSR B1356$-$60,
in Figure 2, shows the descattered pulse profiles and PPA curve observed at 0.659594 $GH_{Z}$; the intrinsic pulse profile is observed at 1.56 $GH_{Z}$; all three models are tested and they produce similar results, the application of thick screen model is presented here. Recovered intensity pulse profiles (c), (d) and PPA curve (g) are similar to the characters of intrinsic pulse signal.

For PSR B1831$-$03,
in Figure 3, shows the descattered pulse profiles and PPA curve observed at 0.4 $GH_{Z}$; the intrinsic pulse profile is observed at 0.61 $GH_{Z}$; all three models has been tested, they produce similar results with different scattering time scales; the application of thin screen model is given here; total intensity and linear intensity profiles match well with those of intrinsic pulse profiles, the recovering of a PPA curve results in a jump-like feature in itself.

For PSR B1838$-$04,
Figure 4 has demonstrated a recovered pulse intensity profiles and PPA curve observed at 0.606 $G_{HZ}$; the intrinsic pulse profile is observed at 0.925 $GH_{Z}$; all three models has been checked; extended and thick screen models produce the same results in intensity profiles and PPA curves, they produced flat PPA curves; here the thin screen model has been applied. Recovered intensity profiles do not match very well with the intrinsic profile, but the PPA curve shows S-curve-like feature (\cite{lorimer05}). In this pulsar another high frequency profile observed at 1.4 $GH_{Z}$ was tried for comparison with recovered pulse profile, but results were not good.
 \begin{figure}

   \centering
   \includegraphics[width=4.5cm, angle=-90]{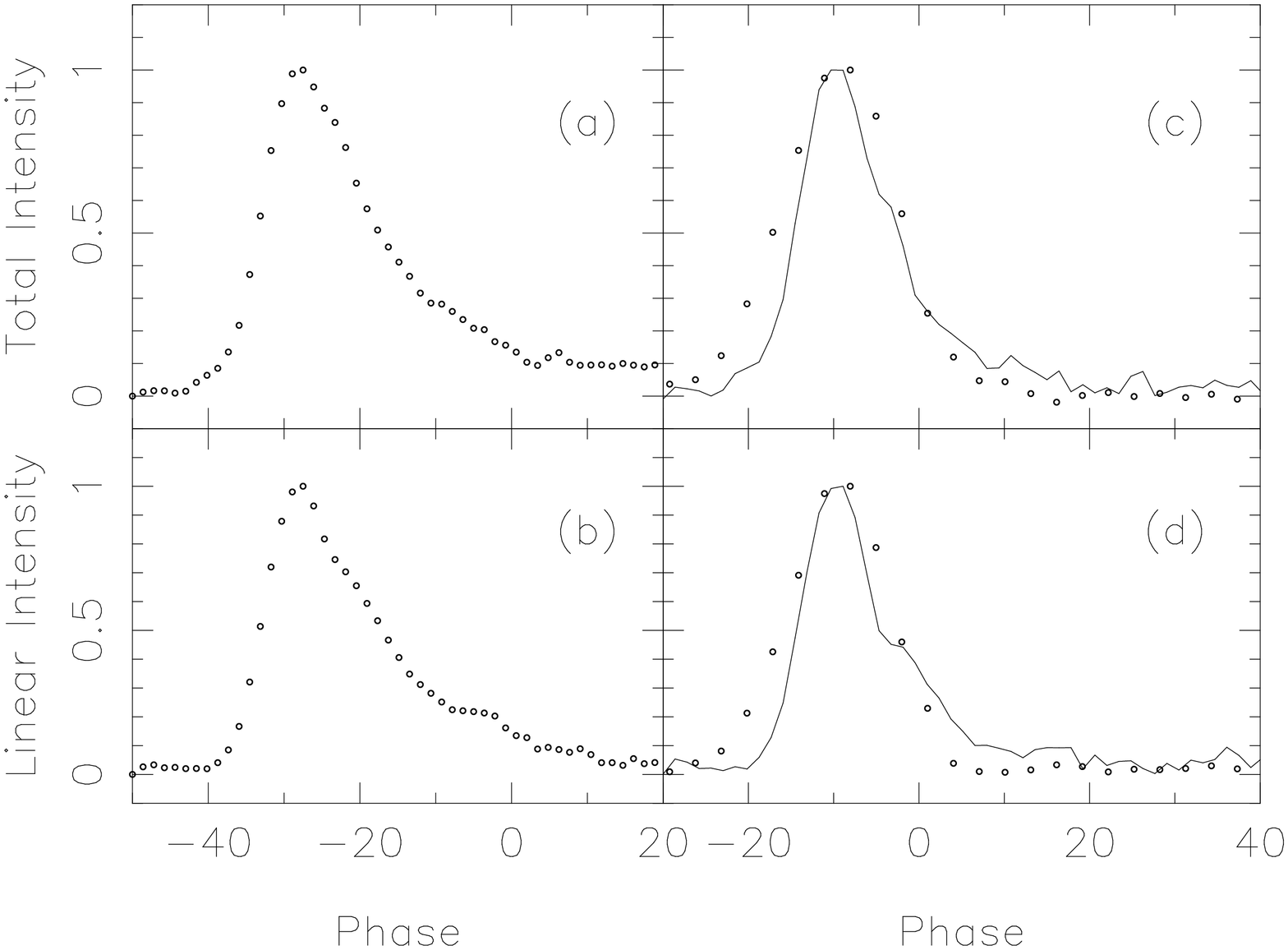}
   \includegraphics[width=4.50cm, angle=-90]{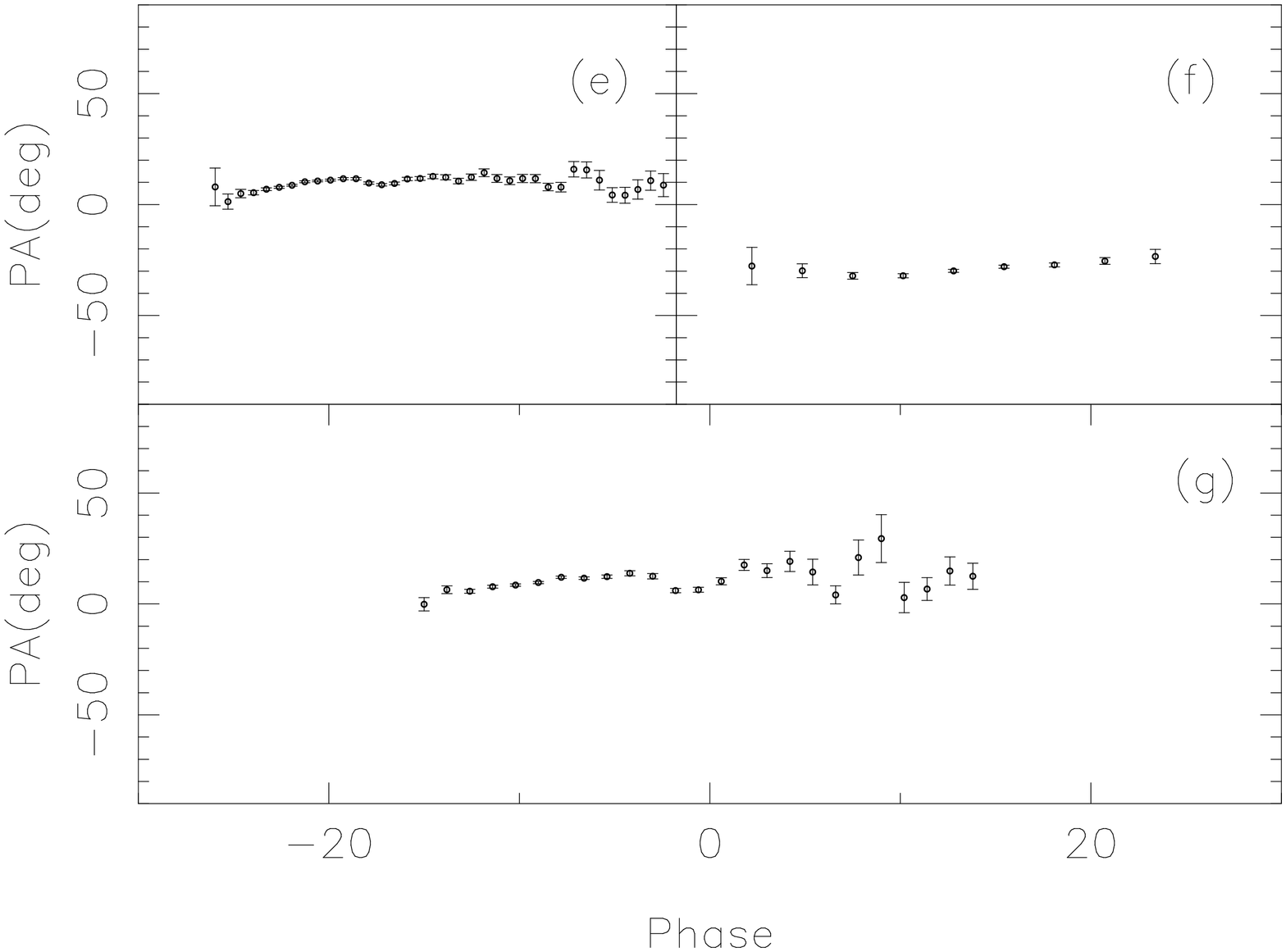}
   \begin{minipage}[]{130mm}
   \caption{ PSR B1356$-$60, the dotted lines in (a), (b) are scattered intensity profiles, in plots (c), (d) solid lines are descattered intensity profiles and the dotted lines are intrinsic intensity profiles for comparison, the plots of (e), (f), (g) are scattered, intrinsic and descattered PPA curves respectively.} \end{minipage}

   \end{figure}

   \begin{figure}
   \centering
   \includegraphics[width=4.50cm, angle=-90]{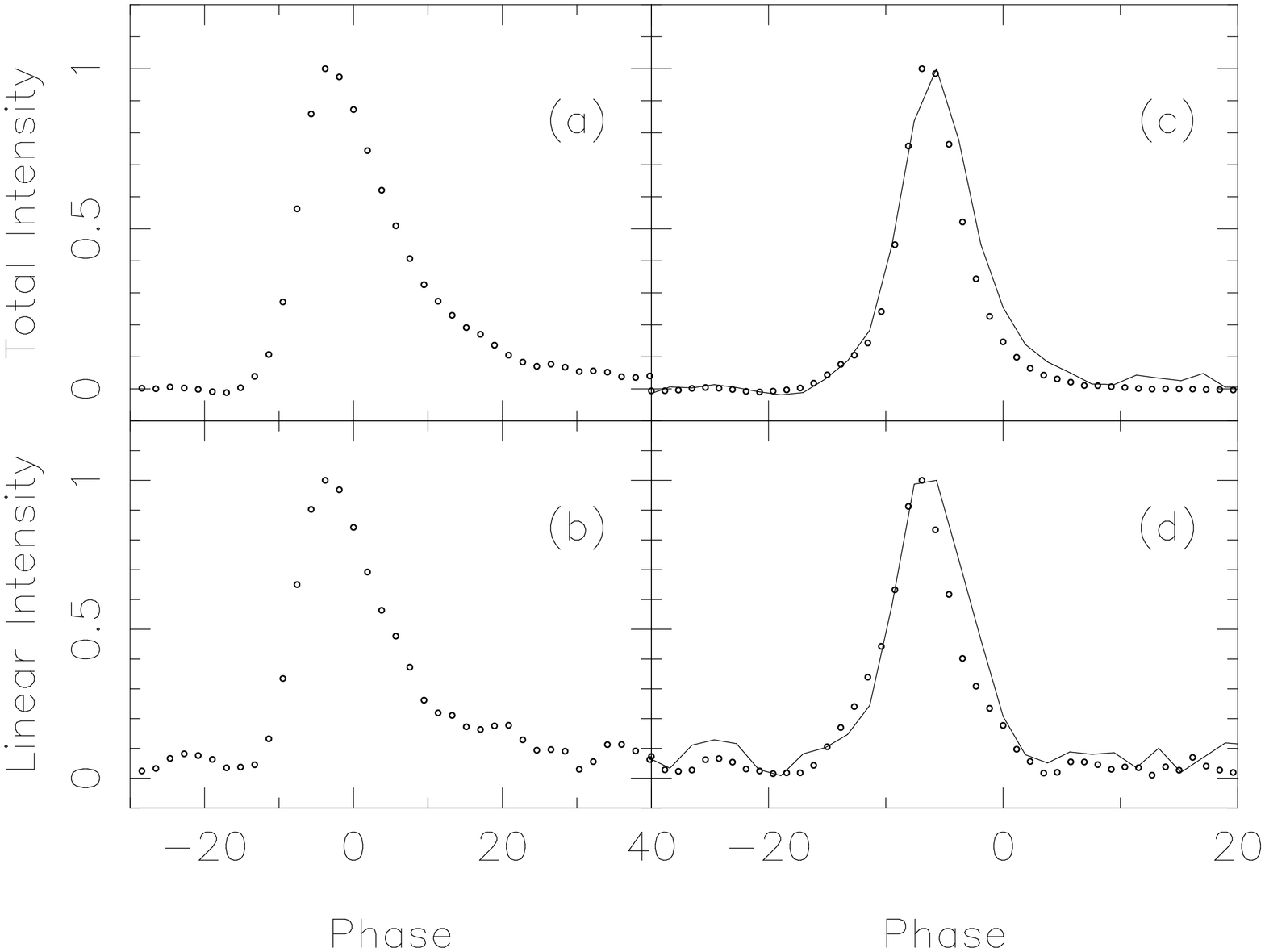}
   \includegraphics[width=4.50cm, angle=-90]{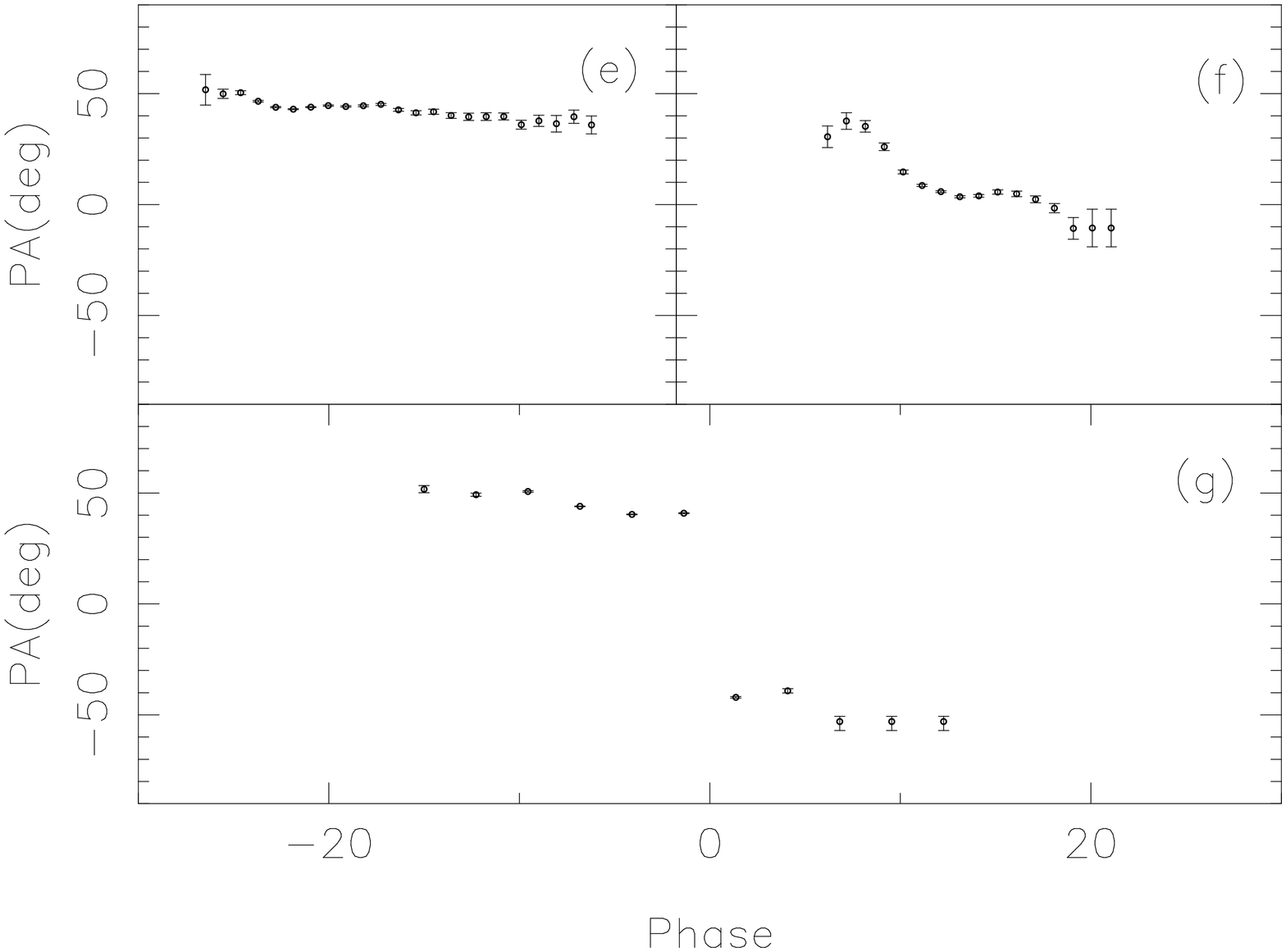}
   \begin{minipage}[]{130mm}

   \caption{ PSR B1831$-$03, the dotted lines in (a), (b) are scattered intensity profiles, in plots (c), (d) solid lines are descattered intensity profiles and the dotted lines are intrinsic intensity profiles for comparison, the plots of (e), (f), (g) are scattered, intrinsic and descattered PPA curves respectively. } \end{minipage}

   \end{figure}

\begin{figure}[!htb]
   \centering
   \includegraphics[width=4.50cm, angle=-90]{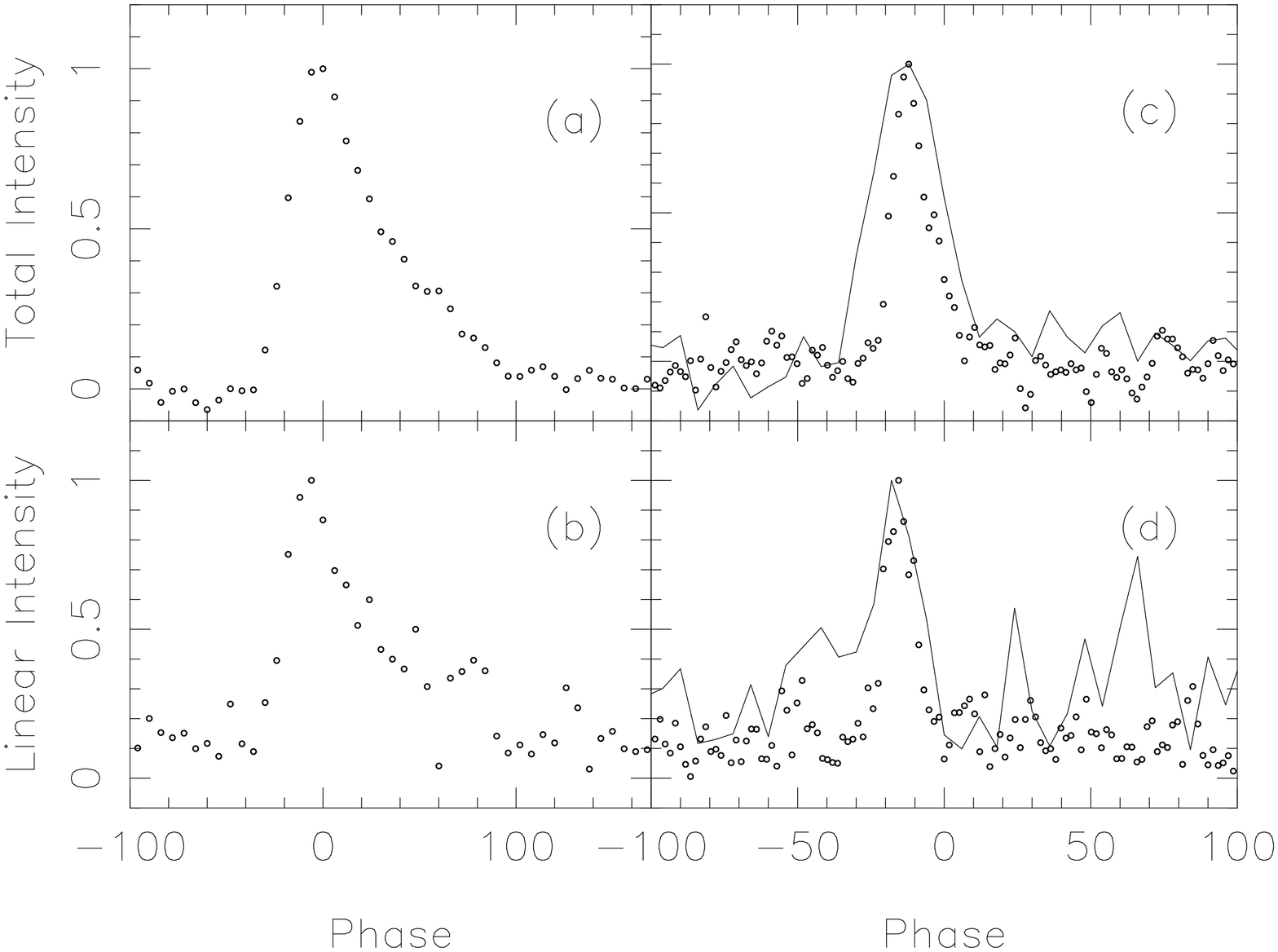}
   \includegraphics[width=4.50cm, angle=-90]{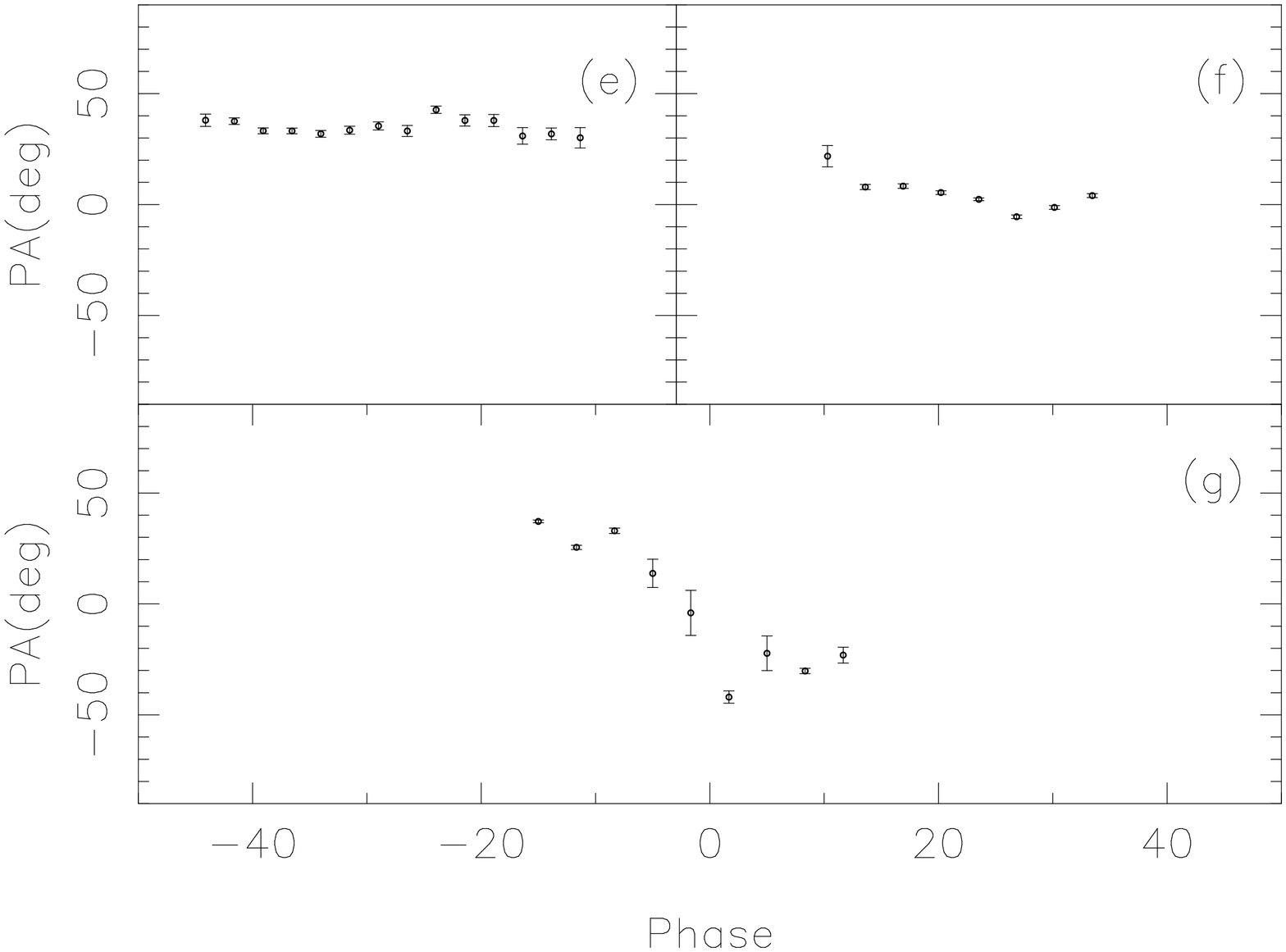}
   \begin{minipage}[]{130mm}

   \caption{ PSR B1838$-$04, the dotted lines in (a), (b) are scattered intensity profiles, in plots (c), (d) solid lines are descattered intensity profiles and the dotted lines are intrinsic intensity profiles for comparison, the plots of (e), (f), (g) are scattered, intrinsic and descattered PPA curves respectively. } \end{minipage}

   \end{figure}

For PSR B1859$+$03,
 Figure 5 plots the descattered pulse profiles and PPA curve observed at 0.606 $GH_{Z}$; the intrinsic pulse profile is observed at 0.925 $GH_{Z}$; all three models are tried, and all of them worked well; the intensity profiles and PPA curves all have the same features. the presented plots in Fig.5 are obtained by using thin screen model. Recovered intensity profiles are similar with the intrinsic profiles in comparison, and PPA curve (g) is much the same with the PPA curve (f) of intrinsic one.

For PSR B1946$+$35,
 Figure 6 gives the descattered pulse profiles and PPA curve observed at 0.408 $GH_{Z}$; the intrinsic pulse profile is observed at 0.61 $GH_{Z}$; all three models are tested, all that models give good results. In extended screen model PPA curve shows flat pattern, in the other two models PPA curves show jump-like structure. Shown here is the application of the thick screen model. Recovered pulse profiles are quite similar to the intrinsic pulse profiles, the recovered PPA curve shows a jump-like feature as shown in Fig. 6. It is acceptable to add 90 degrees to the last four points of curve (g) (\cite{Phrudth09}), making the PPA curve much like with the positive part of plot (f). Interestingly, the higher frequency observation of 0.925 $GH_{Z}$, 1.408 $GH_{Z}$ shows orthogonal jumps in it's PPA curve (\cite{Lorimer98}), this indicates somehow that the recovered PPA curve (g) may be plausible; a fuller discussion will be presented in the next section.

\begin{figure}
   \centering
   \includegraphics[width=4.50cm, angle=-90]{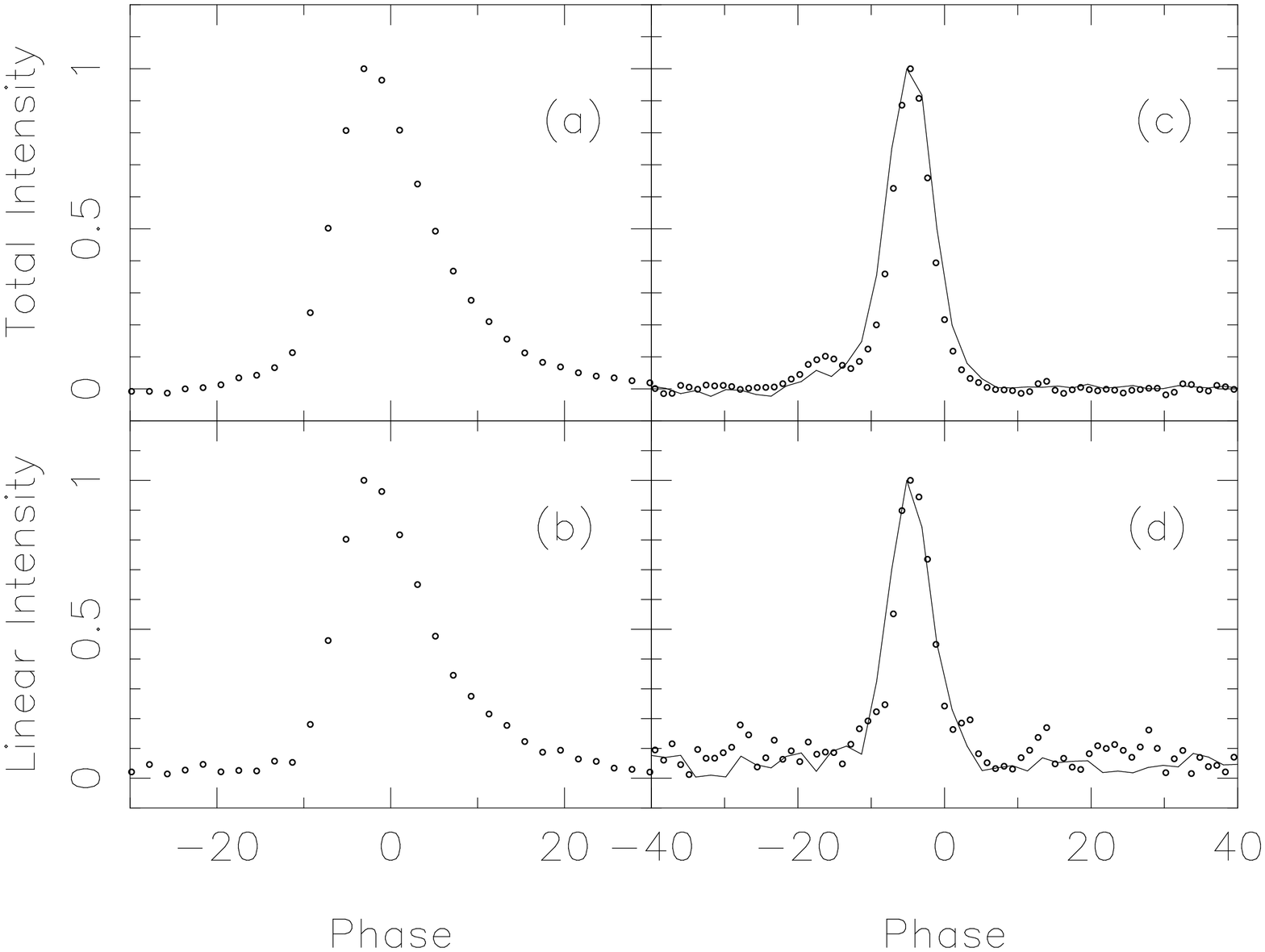}
   \includegraphics[width=4.50cm, angle=-90]{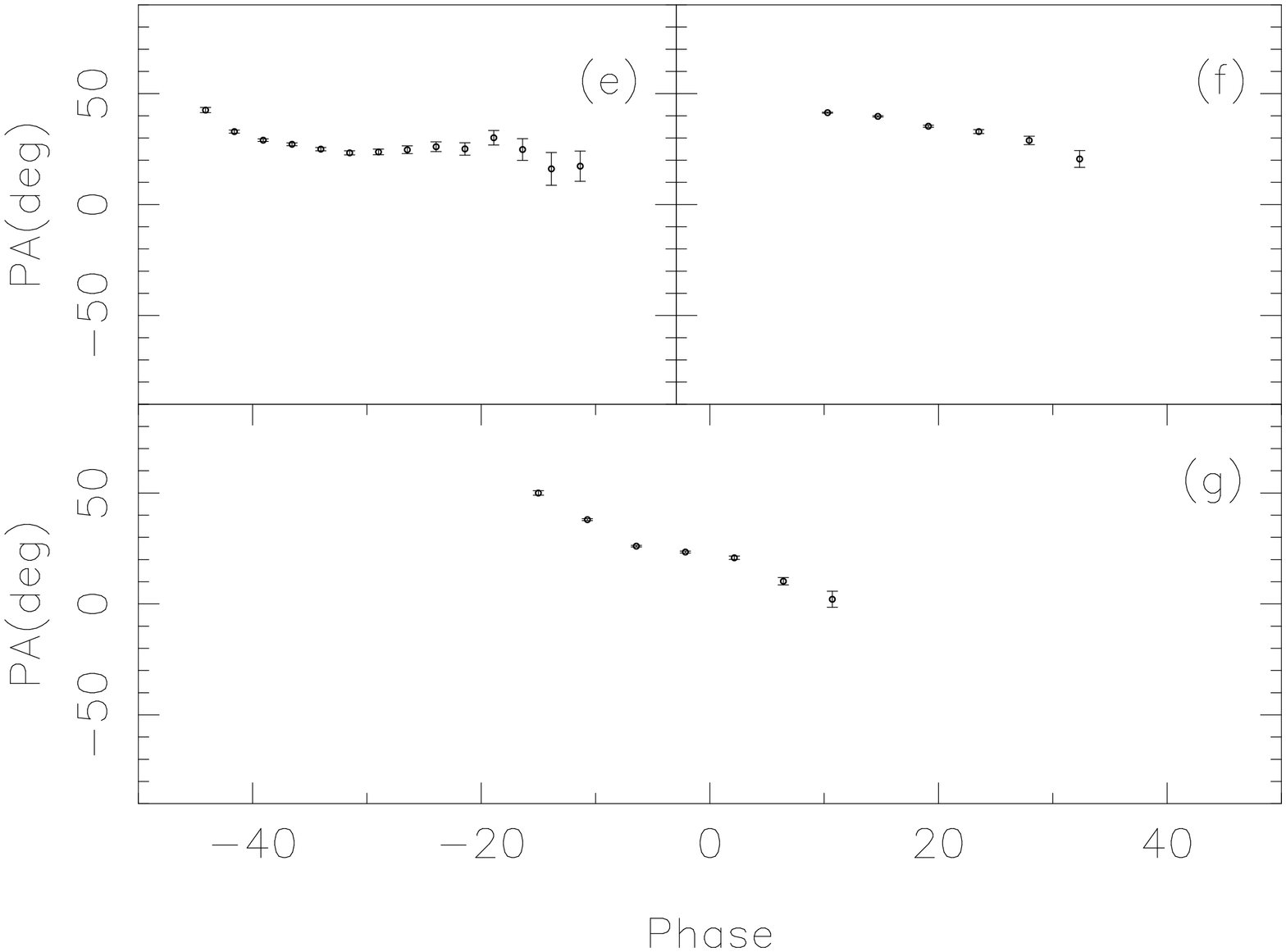}
   \begin{minipage}[]{130mm}

   \caption{ PSR B1859$+$03, the dotted lines in (a), (b) are scattered intensity profiles, in plots (c), (d) solid lines are descattered intensity profiles and the dotted lines are intrinsic intensity profiles for comparison, the plots of (e), (f), (g) are scattered, intrinsic and descattered PPA curves respectively. } \end{minipage}

   \end{figure}

\begin{figure}
   \centering
   \includegraphics[width=4.50cm, angle=-90]{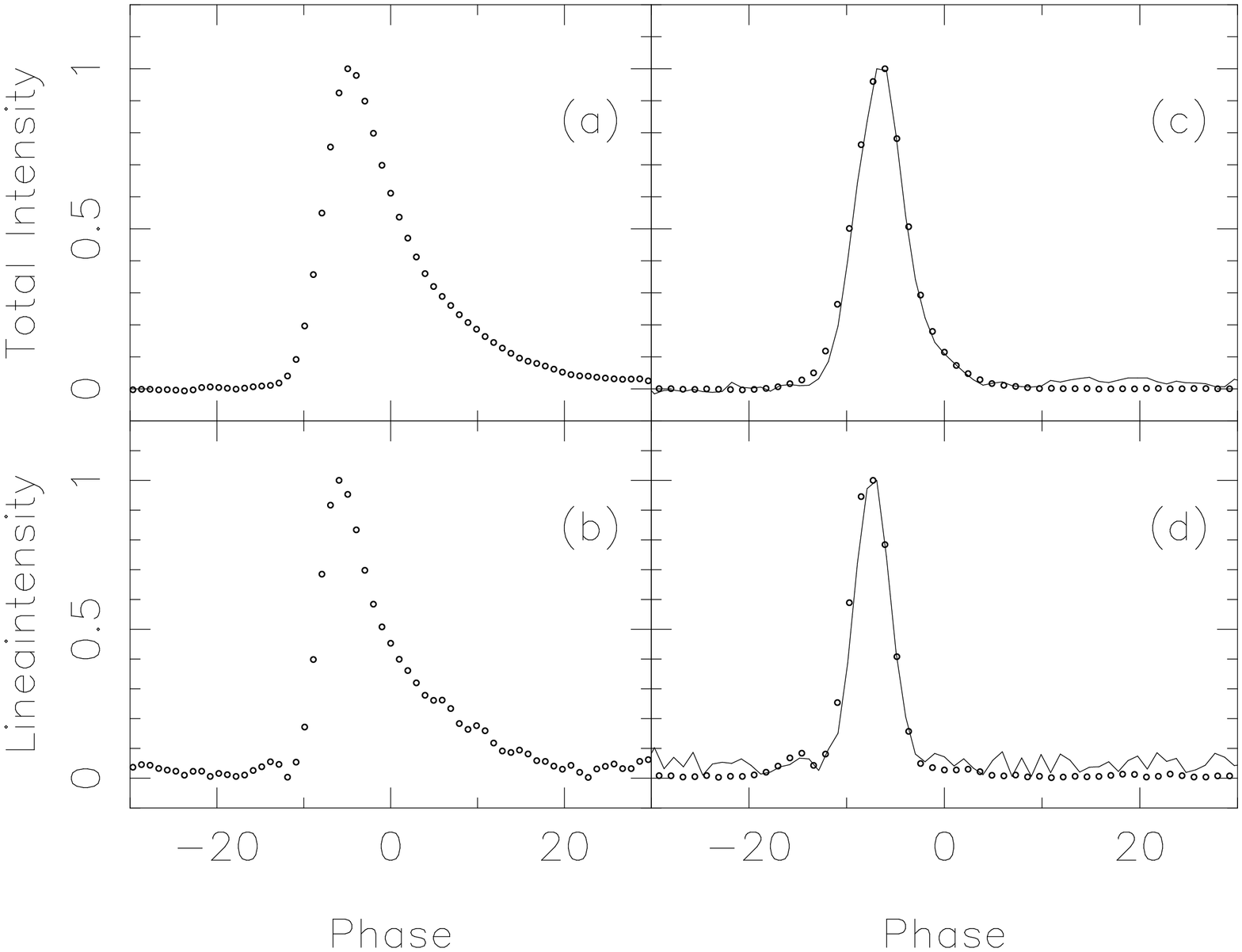}
   \includegraphics[width=4.50cm, angle=-90]{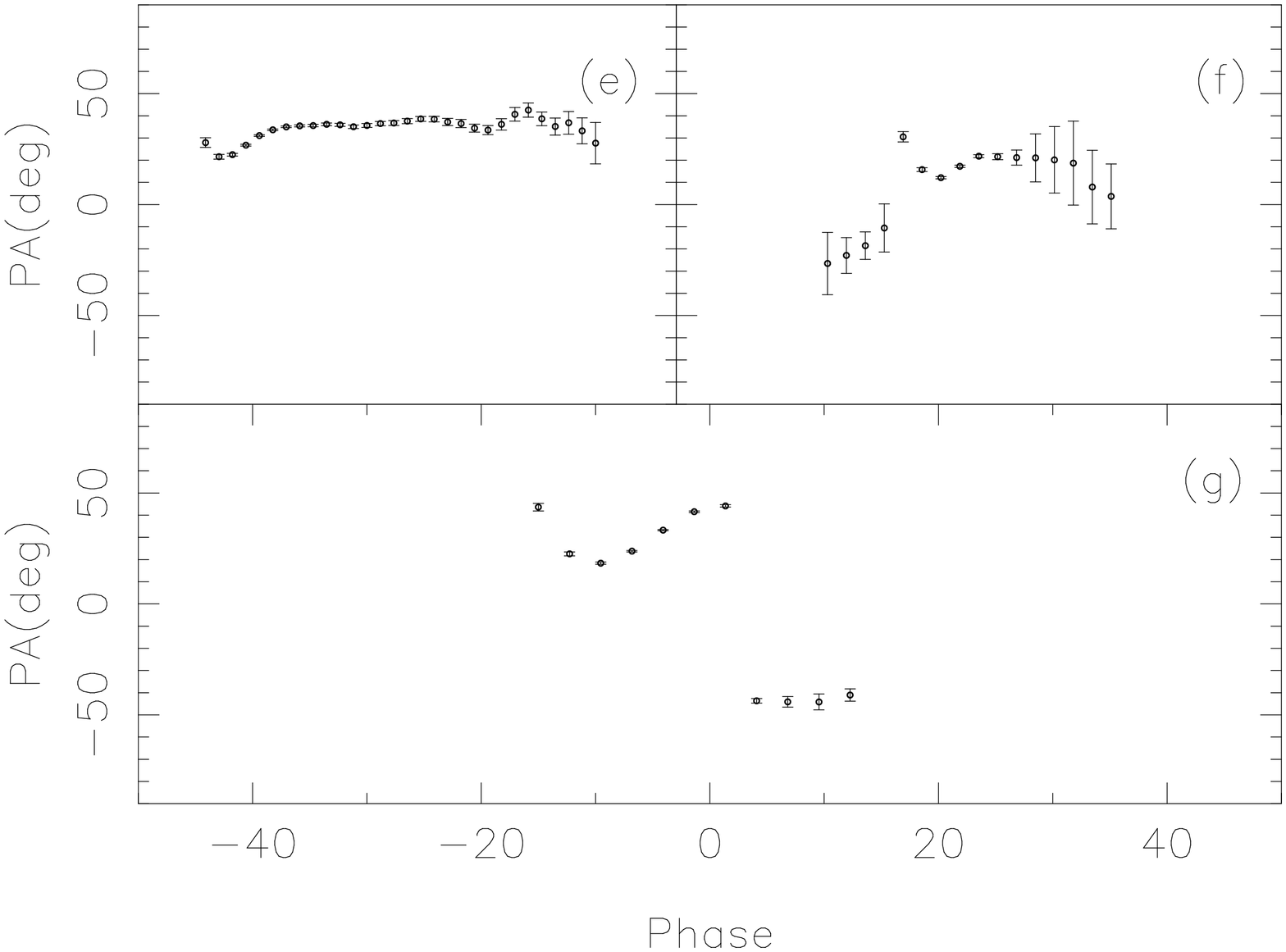}
   \begin{minipage}[]{130mm}

   \caption{ PSR B1946$+$35, the dotted lines in (a), (b) are scattered intensity profiles, in plots (c), (d) solid lines are descattered intensity profiles and the dotted lines are intrinsic intensity profiles for comparison, the plots of (e), (f), (g) are scattered, intrinsic and descattered PPA curves respectively. } \end{minipage}

   \end{figure}








\section{Discussion}
\label{sect:discussion}
 Section 3.1 has shown the simulation of scattering and descattering results; as indicated in Fig. 1 the scattering can cause pulse broadening and flattening of PPA curve; when there is jump in PPA in original pulse, after scattering PPA curve can be much more complicated, but the PPA curves can be descattered as shown in plot (f), (l).  In section 3.2, devoted to practical application, five pulsars' intensity pulse profiles and PPA curves have been descattered; in each pulsar the frequency of intrinsic pulse profile for comparison is below 1.4 $GH{_Z}$. In almost all pulsars the recovered profiles and PPA curves are quite similar to the features of intrinsic pulse profiles and its PPA curves (see Fig. 2, Fig.3, Fig.5, and Fig. 6); these obtained results in section 3.2 support the previous assumption that the original pulse characteristics are substantially frequency-invariant below 1.4 $GH{_Z}$ (\cite{Radhakrishnan69}).

When carrying out the descattering process the existence of pulse width evolution following changes in frequency has been ignored (\cite{lorimer05}) because of the small difference in frequency between the scattered pulse profiles and the intrinsic pulse profiles. Figures (2$-$6) show that all the descattered compensations to the scattered pulse profiles are good except the intensity pulse profiles of PSR B1838$-$04 (see Fig. 4); this may arise from the roughness of scattered pulse profiles. In PSR B1356$-$60, the frequency of a chosen intrinsic pulse profile for comparison is 1.56 $GH_{Z}$, because no other observed frequencies are available below 1.4 $GH{_Z}$. The recovered PPA curves in all pulsars also showed good agreement with our expectation, namely that some of them have similar features to their intrinsic PPA curve (see Figure 2, 5), the PPA curve in Fig. 4 shows S-curve-like feature, some of them have jump-like features (see Figure 3 , 6); the jumps in Figure 3 , 6 can be understood through the simulation in section 3.1; if the original PPA curve has jump feature which were distorted or flattened by scattering (\cite{Karastergiou09}), the recovering of such a PPA curve is likely the cause of jumps within a true scattering model. Fortunately, in higher frequency observation of 0.9 $G H_{Z}$, 1.4 $G H_{Z}$ of the pulsar B1946$+$35 showed orthogonal jumps in their PPA curves (\cite{Lorimer98}); these are the intrinsic pulse profiles compared to the scattered pulse profile observed at 0.408 $GH_{Z}$. The jumps observed are much more likely the intrinsic character of the PPA curves of this pulsar's signal which can be observed below 1.4 $GH_{Z}$. According to the simulation, observational evidence and the empirical assumption of frequency invariance of pulse characters below 1.4 $GH_{Z}$ (\cite{Radhakrishnan69}) it can be said that the occurrence of jumps in our descattering compensation of the PSR B1946$+$35 is acceptable and it would be intrinsic feature of the scattered PPA curve observed at 0.408 $GH_{Z}$. For pulsar B1831$-$3 there is not any observational result of PPA curve with orthogonal jump in EPN database, so it may be easy to explain when adding 90 degrees to the last five points.

\section{Conclusions}
\label{sect:conclusion}

We have shown the descattering compensation to the pulse profiles and PPA curves of five pulsars; the compensation for scattered pulse profiles and PPA curves brought us good results. through simulation and practical application, it is found that all the intrinsic characteristics of pulse signal can be recovered if the scattering model is clear enough. The recovering of pulse characters are an important issue in pulse study and rotation measurement (RM), and in setting the arrival time of impulse signals from pulsar; the recovered S-curve-like PPA curves of such a pulsar B1838$-$04 would give us an opportunity to apply the RVM (\cite{Radhakrishnan69}) to set the  magnetic inclination angle $\alpha$ and impact parameter $\beta$. In one word, the recovering of the pulse profiles and PPA curves may improve our understanding of pulse emission regions and emission mechanisms. In this paper three Stokes parameters $I, Q, U$ have been restored. The Stokes parameter $V$ has been left for latter discussion. The author hopes to continue studying the Stokes phase portraits (\cite{Chung10}) of those five pulsars which may be useful for the research in pulsar emission properties.

\normalem
\begin{acknowledgements}
We thank the referee for the helpful comments. This work was funded by the National Natural Science Foundation of China (NSFC) under No.10973026 and the key program project of Joint Fund of Astronomy by NSFC and CAS under No. 11178001.

\end{acknowledgements}



\label{lastpage}

\end{document}